\documentclass[lettersize,journal]{IEEEtran}
\usepackage{amsmath,amsfonts}
\usepackage{algorithmic}
\usepackage{algorithm}
\usepackage{array}
\usepackage{subcaption}
\usepackage{textcomp}
\usepackage{stfloats}
\usepackage{url}
\usepackage{verbatim}
\usepackage{graphicx}
\usepackage{cite}
\usepackage{booktabs}
\usepackage{multirow}
\usepackage{colortbl}
\usepackage[table]{xcolor}
\usepackage{hyperref} 
\definecolor{bluecol}{rgb}{0.85,0.9,1}
\definecolor{orangecol}{rgb}{1,0.9,0.8}
\definecolor{greencol}{rgb}{0.85,1,0.85}
\definecolor{headerblue}{rgb}{0.2,0.4,0.6}
\definecolor{proposed}{rgb}{0.9,0.9,0.9}

\begin{document}

\title{PRISM: Decision-Centric Predictive Sensing for Cognitive Digital Twins in 6G}

\author{Afan Ali,~\IEEEmembership{Member,~IEEE}, Daniel Benevides da Costa,~\IEEEmembership{Senior Member,~IEEE}, and \\Ali Arshad Nasir,~\IEEEmembership{Senior Member,~IEEE}
\thanks{The authors are with the Interdisciplinary Research Center for Communication Systems and Sensing (IRC-CSS), Department of Electrical Engineering, King Fahd University of Petroleum and Minerals (KFUPM), Dhahran 31261, Saudi Arabia.}}

\markboth{Journal of \LaTeX\ Class Files,~Vol.~, No.~, May~2026}%
{Shell \MakeLowercase{\textit{et al.}}: A Sample Article Using IEEEtran.cls for IEEE Journals}

\IEEEpubid{}

\maketitle

\vspace{-3cm}

\begin{abstract}
Integrated sensing and communication (ISAC) and Digital Twin (DT) technology have emerged as complementary for future wireless networks that require autonomous operations involving continuous interaction between physical and digital worlds. However, existing DT-assisted ISAC frameworks sense continuously and indiscriminately while optimizing only a single task, leaving little room for persistent, multi-domain knowledge or proactive sensing control. This article proposes a Predictive, Reasoning-driven, Intelligent Sensing Module (PRISM) engine that transforms the DT from a passive, domain-specific optimizer into a persistent, network-wide reasoning system. PRISM enables decision-centric predictive perception, proactively directing sensing toward anticipated decisions needs rather than following fixed sensing schedules. Using an illustrative extremely large multiple-input multiple-output (XL-MIMO) deployment scenario with a mixed eMBB, URLLC, and mMTC device population, we show how this principle benefits visibility-region sensing for channel acquisition and supports slice-aware operation. Preliminary simulations, including this deployment scenario and the resulting knowledge error, overhead, and latency results, confirm that this decision-centric approach substantially reduces sensing overhead while preserving decision reliability and latency, supporting the proposed architecture as a practical step toward self-aware, autonomously orchestrated 6G networks.
\end{abstract}

\section{Introduction}

\begin{figure}[t]
\centering

    \centering
    \fbox{\includegraphics[width=0.96\linewidth,
                           height=4.7cm,
                           keepaspectratio]{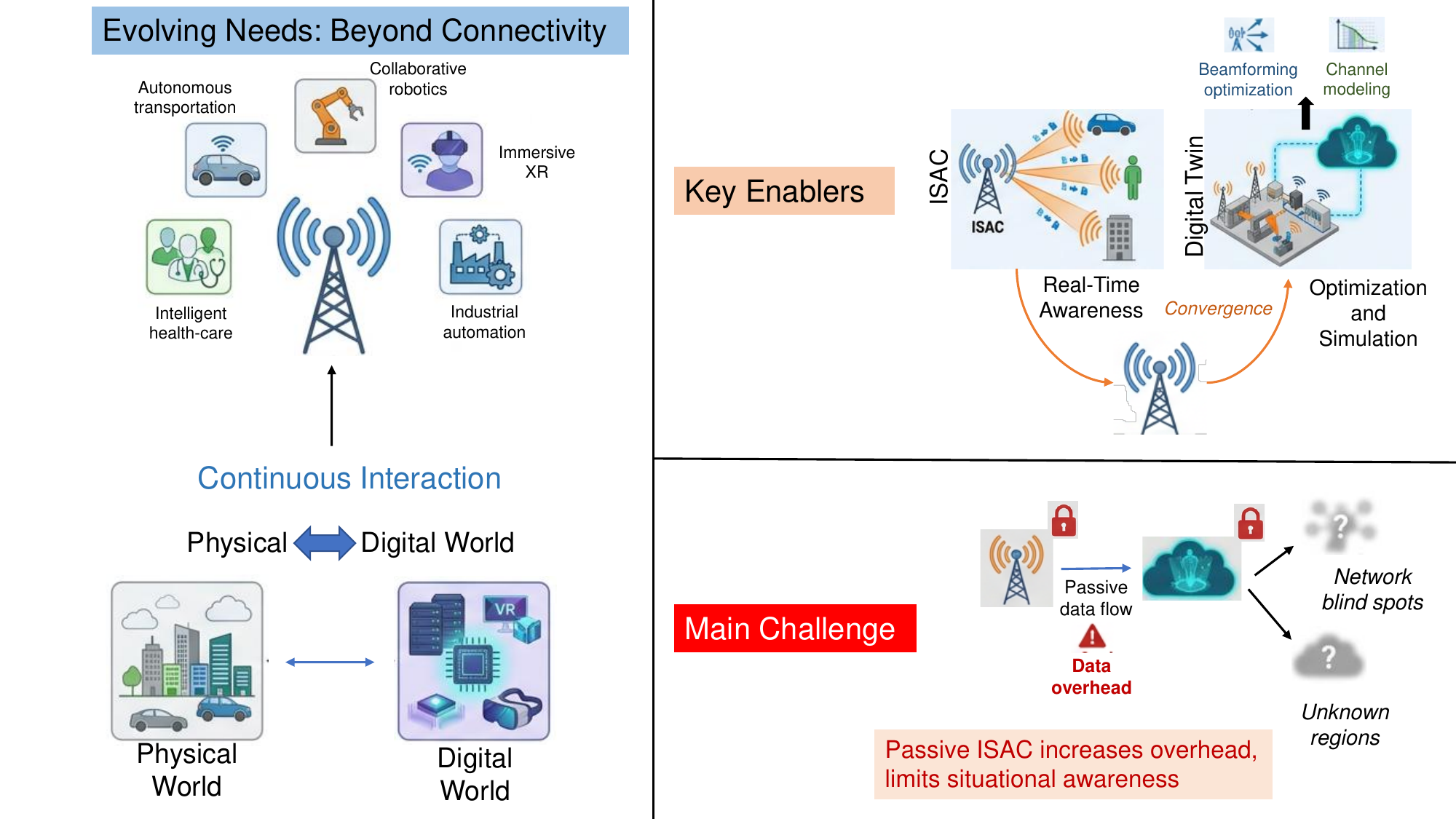}}
    \caption{Emerging 6G requirements, the convergence of ISAC and digital twins as key enablers, and the challenges introduced by passive ISAC.}
\label{fig1}
\end{figure}

\IEEEPARstart{S}{ixth-generation} (6G) wireless networks are expected to move beyond connectivity to support autonomous transportation, industrial automation, immersive extended reality (XR), collaborative robotics, and large-scale cyber-physical systems, all of which demand continuous interaction between the physical and digital worlds. Realizing this vision requires bringing perception, communication, computing, and intelligence together into a single operational framework~\cite{wei_integrated_2024}. As depicted in Fig.~\ref{fig1}, two technologies have emerged as central enablers of this shift. Integrated sensing and communication (ISAC) equips wireless infrastructure with the ability to jointly communicate and perceive its surroundings using shared radio resources, enabling real-time localization, environmental mapping, and situational awareness~\cite{zhang_integrated_2025}, while Digital Twin (DT) technology provides continuously synchronized virtual representations that let operators monitor, predict, and validate network behavior before decisions are deployed physically. Their complementary strengths have driven rapid convergence of ISAC and DT as foundational building blocks for AI-native 6G.

Table~\ref{tab:comparison} summarizes representative developments along this direction, spanning an ISAC-driven DT architecture for intelligent machine networks~\cite{wei_integrated_2024}, a DT-assisted ISAC framework for predictive beamforming and vehicle association~\cite{ding_joint_2024}, a Radio Digital Twin (RDT) that fuses ISAC, computer vision, and wireless measurements for beam prediction~\cite{dai_scalable_2025}, Holographic Digital Twins (HDT) for immersive communication~\cite{zhang_integrated_2025}, DT migration for adaptive edge service provisioning~\cite{ma_adadt_2026}, and generative channel twins for statistical channel state information (CSI) generation, beam prediction or channel estimation~\cite{dong_accelerated_2026,zhang_efficient_2026,wang_digital_2026}. A two-step goal-oriented access scheme is proposed in~\cite{saggese_goal-oriented_2026} in which high-value of information (VoI) sensors request resources via push-based random access before pull-based transmissions deliver ISAC sensory data to the DT, maximizing delivered VoI while localizing the sensors. Despite spanning different application domains, these works share a common architectural philosophy, i.e, ISAC acts as a sensing mechanism that keeps the virtual representation updated, while the DT is used to optimize one specific wireless task, such as beamforming, channel estimation, mobility management, or resource allocation. This leaves the DT behaving more like an application-specific optimization platform than a persistent knowledge system for autonomous operation, and little attention has gone into how a DT should interpret observations, retain knowledge across domains, or proactively guide future sensing. This architecture also leaves telecom operators outside the loop: existing frameworks treat operator objectives and service-level policies as external inputs applied after optimization rather than as signals that actively shape sensing and decision-making, so operators can neither inject intent in real time nor receive feedback on why the network chose a given action. This limits trust, auditability, and the network's ability to align its behavior with commercial or regulatory priorities. The problem is compounded by the largely passive role ISAC continues to play throughout this literature, sensing observes the environment continuously and indiscriminately regardless of what the DT actually needs to know, inflating overhead while limiting the network's ability to focus resources on the regions that matter most. This gap is set to widen as 6G networks adopt extremely large multiple-input multiple-output (XL-MIMO) transmission and split their resources across network slices, such as, enhanced mobile broadband (eMBB), ultra-reliable low-latency communication (URLLC), and massive machine-type communication (mMTC), each with very different latency and bandwidth budgets. These requirements multiply the amount of state a passive DT would need to track.

Motivated by this gap, this article proposes a new architectural framework for AI-native 6G in which ISAC, DTs, and AI operate as tightly coupled components of a continuous loop of perception, reasoning, action, and learning. The framework reshapes the DT from a passive, domain-specific optimizer into a persistent, network-wide Predictive, Reasoning-driven, Intelligent Sensing Module (PRISM) engine, built around a design principle we call decision-centric predictive perception, in which sensing is directed by anticipated decisions rather than driven on a fixed schedule. PRISM also addresses the operator-in-the-loop gap identified above by incorporating operator objectives directly into the reasoning process as first-class inputs, rather than treating them as post-hoc constraints. Moreover, PRISM establishes a bidirectional feedback loop in which decisions and their outcomes are communicated back to the operator, while sensing information flows to the DT, contrasting with existing pipelines that primarily rely on one-directional optimization. The main contributions of this article are as follows:

\begin{itemize}
\item \textbf{Decision-centric predictive perception:} sensing is proactively guided by anticipated decisions rather than a predetermined schedules.
\item \textbf{Persistent multi-domain cognitive memory:} knowledge is persistently retained and reused across communication, sensing, mobility, computing, and services to support cross-domain reasoning.
\item \textbf{Anticipatory orchestration:} reasoning predicts future decisions, and proactively planning pre-stages candidate actions, enabling timely orchestration before those decisions arise.
\item \textbf{Operator-in-the-loop bidirectional feedback:} operator objectives directly guide reasoning process, while decisions and their outcomes are fed back for operator visibility.
\end{itemize}

\begin{table*}[!t]
\caption{Comparison of representative DT-assisted ISAC frameworks and their architectural scope.
$\checkmark$ = Fully supported, $\triangle$ = Partially supported, $\times$ = Not considered.}
\label{tab:comparison}
\centering
\scriptsize
\renewcommand{\arraystretch}{1.05}
\begin{tabular}{
|>{\centering\arraybackslash}p{1cm}
|>{\columncolor{bluecol}}p{5.4cm}
|>{\centering\arraybackslash\columncolor{orangecol}}p{1.5cm}
|>{\centering\arraybackslash\columncolor{orangecol}}p{1.5cm}
|>{\centering\arraybackslash\columncolor{greencol}}p{1.5cm}
|>{\centering\arraybackslash\columncolor{greencol}}p{1.5cm}|
}
\hline
\rowcolor{headerblue}
\textcolor{white}{\textbf{Ref.}} &
\textcolor{white}{\textbf{DT Scope and Function}} &
\textcolor{white}{\textbf{Goal-\newline Oriented}} &
\textcolor{white}{\textbf{Bidirectional\newline Feedback}} &
\textcolor{white}{\textbf{Network-wide\newline Cognition}} &
\textcolor{white}{\textbf{Operator-\newline in-Loop}}
\\
\hline
\cite{wei_integrated_2024} & Machine network state; comm./network optimization & $\times$ & $\triangle$ & $\times$ & $\times$ \\
\hline
\cite{ding_joint_2024,yang_near-field_2026} & Vehicle mobility \& semantics; beamforming, resource allocation & $\triangle$ & $\checkmark$ & $\times$ & $\times$ \\
\hline
\cite{dai_scalable_2025,montaner_deterministic_2026} & Radio/RF twin; environment reconstruction, channel modeling & $\times$ & $\triangle$ & $\times$ & $\times$ \\
\hline
\cite{dong_accelerated_2026,wang_digital_2026} & Channel/CSI generation; estimation and prediction & $\times$ & $\times$ & $\times$ & $\times$ \\
\hline
\cite{zhang_integrated_2025} & Holographic scene twin; immersive communication & $\times$ & $\triangle$ & $\times$ & $\times$ \\
\hline
\cite{ma_adadt_2026} & Edge service state; DT migration, orchestration & $\triangle$ & $\triangle$ & $\times$ & $\times$ \\
\hline
\cite{saggese_goal-oriented_2026,luo_electromagnetic_2026} & VoI-driven sync; closed-loop beam management & $\checkmark$ & $\triangle$ & $\times$ & $\times$ \\
\hline
\rowcolor{proposed}
\textbf{PRISM} & \textbf{Persistent multi-domain knowledge (comm., sensing, compute, mobility, services, intent); network-wide orchestration} & $\checkmark$ & $\checkmark$ & $\checkmark$ & $\checkmark$ \\
\hline
\end{tabular}
\end{table*}

\section{Related Work: From Domain-Specific Digital Twins to Network Cognition}
\label{secII}

\subsection{From Domain-Specific Intelligence to Network Cognition}

The concept of the cognitive digital twin (CDT) grew out of the realization that conventional DTs could move beyond passive synchronization toward genuine decision support. Early work focused mainly on manufacturing, where cognition was introduced through capabilities such as perception, memory, reasoning, and learning to improve production efficiency and autonomous operation~\cite{mortlock_graph_2022}. Similar ideas later found their way into healthcare and smart city applications, where CDTs support personalized diagnosis, predictive analytics, and intelligent urban management through continuous data fusion and AI-driven decision making~\cite{zhang_enabling_2026,shiraptini_ai_2026}. With the rise of ISAC and AI-native networking, wireless communications have recently embraced the DT paradigm as well~\cite{wei_integrated_2024,ding_joint_2024,dai_scalable_2025,zhang_integrated_2025,dong_accelerated_2026,wang_digital_2026,ma_adadt_2026}. 

However, existing DTs are almost always built around one specific network entity, such as wireless channels, radio environments, vehicle mobility, electromagnetic propagation, or edge services, and are then used to optimize the corresponding network function. Although these frameworks show increasingly sophisticated intelligence, their knowledge stays confined to a single application domain.

\subsection{Why AI-Native 6G Requires Network Cognition}

Domain-specific DTs have clearly improved individual wireless functions, but AI-native 6G brings requirements that go well beyond optimizing isolated communication tasks. Future wireless networks will operate as highly distributed cyber-physical ecosystems in which communication, sensing, computing, mobility, edge intelligence, and network services continuously interact across multiple spatial and temporal scales. This interaction only grows harder to manage as base stations move to extremely large antenna arrays for XL-MIMO transmission and as operators partition capacity across eMBB, URLLC, and mMTC network slices with sharply different latency and reliability targets, both of which increase the amount of state a DT must track. We approximate this challenge later in Sec.~\ref{secIV} through a lightweight, system-level use case rather than a full physical-layer XL-MIMO or slice-differentiated evaluation, which we leave for future work. Decisions made in one domain inevitably ripple into others. Optimizing these functions in isolation can, therefore, no longer deliver globally efficient network operation.

This growing interdependence also changes the role of both ISAC and the DT. Existing DT-assisted ISAC frameworks typically follow a sequential workflow: ISAC acquires environmental observations, the DT synchronizes its virtual representation, and an optimization algorithm decides the next network action. This has proven effective for application-specific problems such as channel estimation, beam management, mobility prediction, and edge service migration, but the DT remains a passive consumer of information, with little say in what should be sensed next or how knowledge gathered across domains should shape future operation. As Table~\ref{tab:comparison} shows, this knowledge is also typically confined to a single domain, whether wireless channels, radio environments, vehicle mobility, or edge services. These specialized DTs perform their individual tasks well, but each offers only a partial view of the network. AI-native 6G instead calls for a DT that continuously brings together heterogeneous information into one unified representation.

\subsection{Toward a Predictive, Reasoning-driven, Intelligent Sensing Module (PRISM): Our Approach}

\begin{figure*}[t]
\centering
\captionsetup[subfigure]{
    font={footnotesize},
    labelfont=bf,
    justification=centering,
    skip=3pt
}

\begin{subfigure}[b]{0.68\textwidth}
    \centering
    \fbox{\includegraphics[width=1\linewidth,
                           height=4.9cm,
                           keepaspectratio]{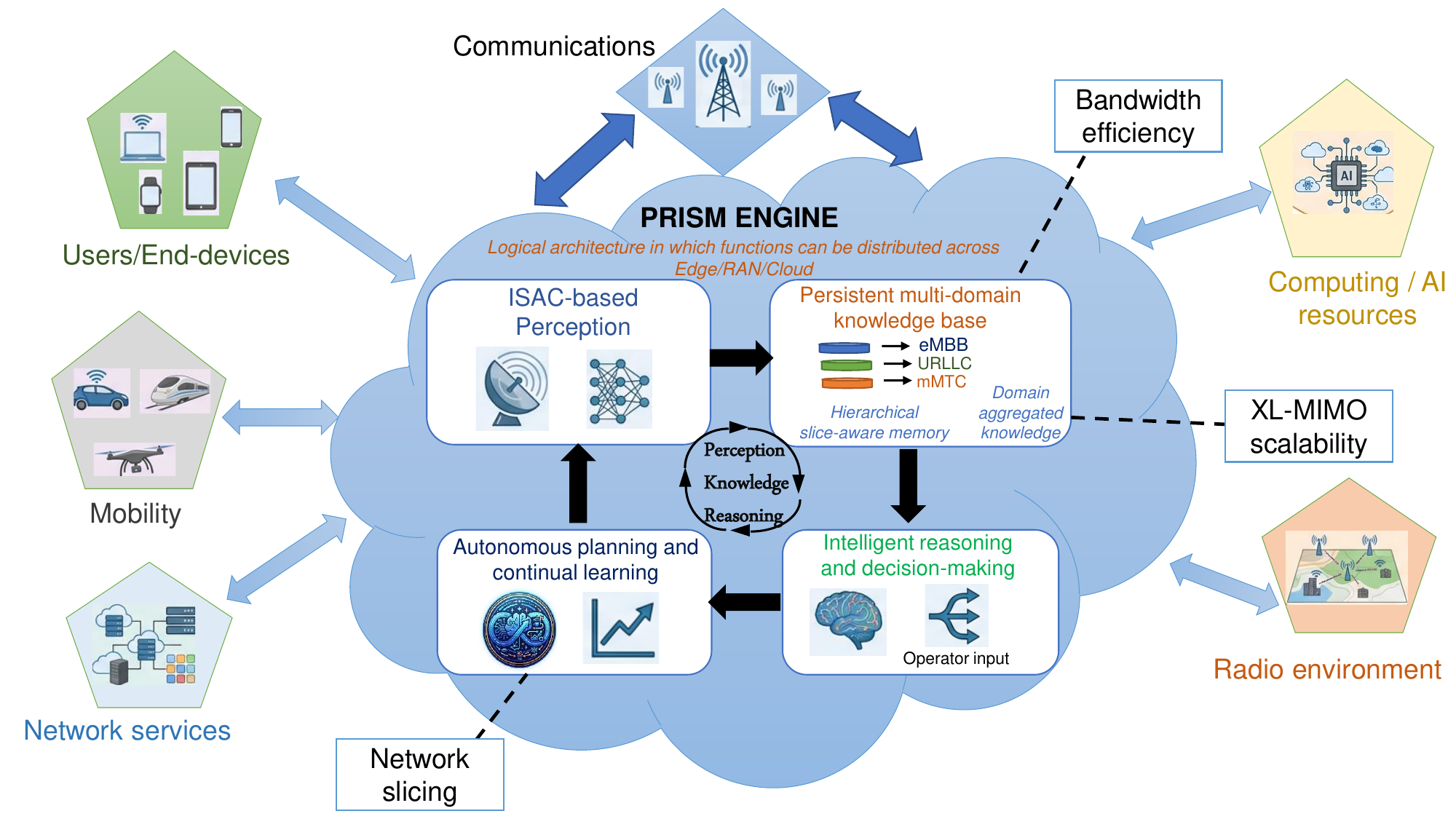}}
    \caption{}
    \label{fig2a}
\end{subfigure}

\vspace{0mm}

\begin{subfigure}[b]{0.48\textwidth}
    \centering
    \fbox{\includegraphics[width=0.98\linewidth,
                           height=4.5cm,
                           keepaspectratio]{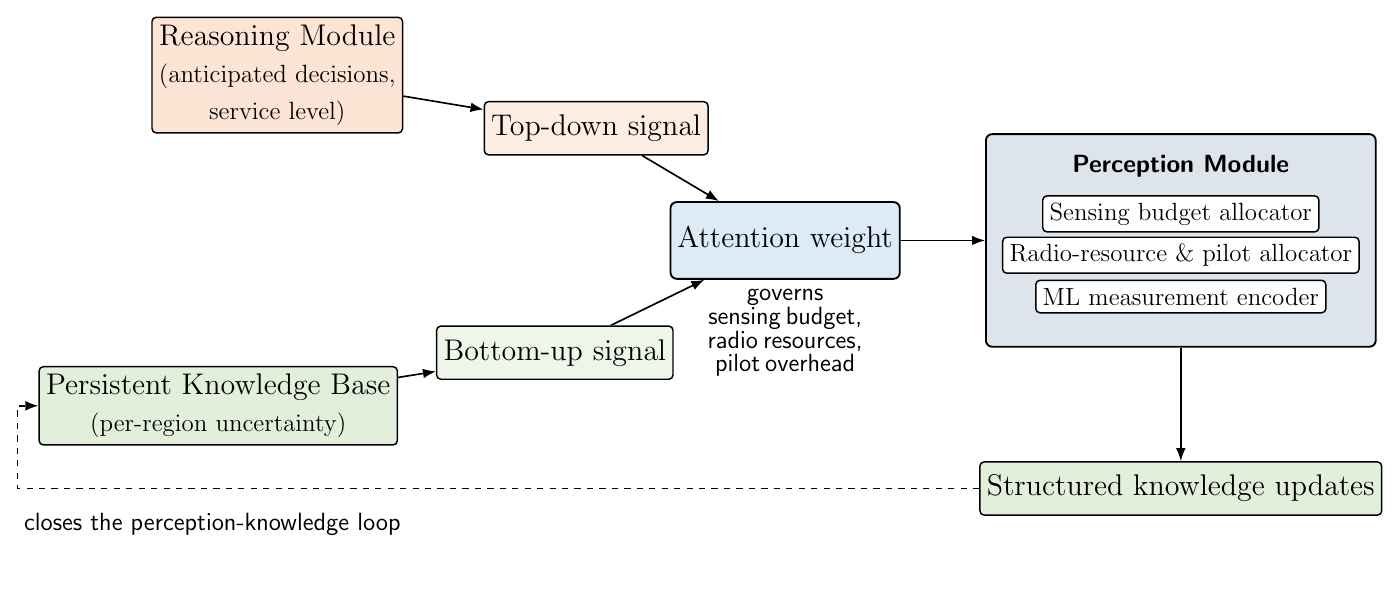}}
    \caption{}
    \label{fig2b}
\end{subfigure}
\hfill
\begin{subfigure}[b]{0.48\textwidth}
    \centering
    \fbox{\includegraphics[width=0.98\linewidth,
                           height=4.5cm,
                           keepaspectratio]{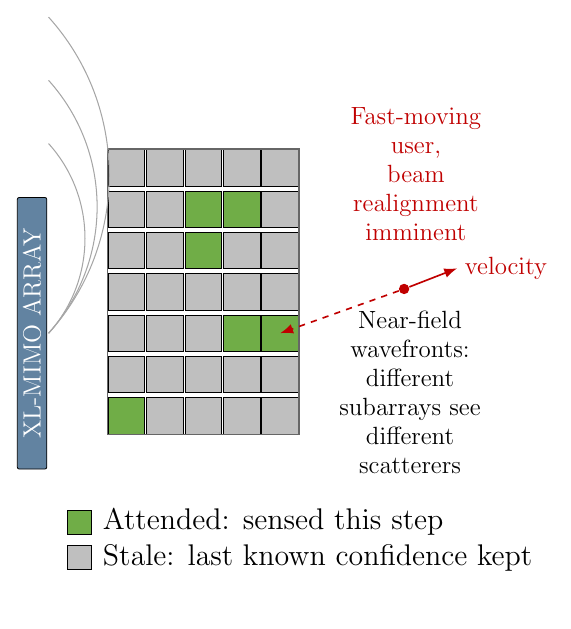}}
    \caption{}
    \label{fig2c}
\end{subfigure}
\vspace{0mm}

\caption{
(a)~PRISM: Predictive, Reasoning-driven, Intelligent Sensing Module engine.
(b)~Bottom-up uncertainty and top-down anticipated-decision signals combine into a single attention weight.
(c)~The XL-MIMO array indexed by visibility region: only regions tied to an active or anticipated decision, such as a fast-moving user needing beam realignment, are resensed.
}

\label{fig2}

\end{figure*}

To close these gaps, this article reshapes the DT from a passive, domain-specific optimizer into a persistent, network-wide PRISM engine, shown in Fig.~\ref{fig2a}. Instead of letting ISAC sense the environment continuously and indiscriminately, the PRISM engine treats perception as goal-driven, directing sensing toward whatever the knowledge base still needs to resolve instead of repeating it regardless of relevance. Rather than relying on isolated, task-specific representations, PRISM integrates communication, user, radio-environment, mobility, and network-resource information into one persistent multi-domain knowledge base, enabling knowledge to be retained and reused across tasks instead of being reconstructed for each task. Built on top of this knowledge base, an intelligent reasoning layer interprets the current network state and weighs competing objectives across domains, while an autonomous planning and continual-learning stage converts these decisions into control actions and feeds their outcomes back into the knowledge base. Perception, knowledge, reasoning, and planning are connected in a closed loop, so the engine keeps refining what it senses, what it knows, and what it does as the network evolves and as operator objectives shift.

\section{Proposed PRISM Architecture}
\label{secIII}

\subsection{Overview}

Fig.~\ref{fig2a} illustrates the PRISM engine, in which it sits at the core of the network and continuously exchanges information with six operational domains: communications, users and end-devices, the radio environment, computing and AI resources, mobility, and network services. Internally, it is organized around four tightly coupled blocks, an ISAC-based perception module, a persistent multi-domain knowledge base, an intelligent reasoning and decision-making module, and an autonomous planning and continual-learning module, arranged around the closed perception-knowledge-reasoning loop introduced in Section~\ref{secII}.

\subsection{Perception: Attention-Guided, Predictive ISAC Sensing}

In conventional DT-assisted ISAC frameworks, perception is a one-way, bottom-up process, i.e., sensing runs on a fixed schedule or in response to raw uncertainty, and whatever it discovers is pushed upward to the rest of the system. The perception module in the PRISM engine instead operates under two combined signals, illustrated in Fig.~\ref{fig2b}. A bottom-up signal still tracks how uncertain each part of the knowledge base currently is, but a second, top-down signal originates in the reasoning module and specifies which parts of the network state actually need refreshing to support the next anticipated decision. The two signals are combined into a single attention weight that governs how the available sensing budget, radio resources, and pilot overhead are distributed across the network at any given moment, closing the loop back to the knowledge base through structured updates. In the present prototype this combination follows a simple priority rule rather than a learned weight, regions tied to an anticipated decision are attended first, and any sensing budget left over is assigned to the regions with the highest bottom-up uncertainty, leaving a learned or continuously tuned weighting for later implementations.

This attention-guided design pays off directly for XL-MIMO deployments, where large antenna apertures introduce near-field propagation and spatial non-stationarity, so different subarrays effectively see different scatterers and users within their own visibility regions, as shown in Fig.~\ref{fig2c}. Estimating the full near-field channel on a fixed schedule is prohibitively expensive in pilot overhead and processing time. The proposed perception module indexes the array by visibility region rather than antenna element, and the attention mechanism refreshes only the regions tied to an active or anticipated decision, for instance the subarrays serving a fast-moving user about to need beam realignment. Regions with no pending decision keep their last known confidence level rather than being re-measured, which keeps near-field acquisition overhead roughly proportional to the number of decisions the network is making rather than to array size. Machine learning models embedded in the perception module translate these sparse, targeted measurements into structured updates the knowledge base can absorb directly, rather than raw signals reprocessed for every downstream task.

\subsection{Persistent Multi-Domain Knowledge Base: Hierarchical and Slice-Aware}

At the center of the PRISM engine sits a persistent multi-domain knowledge base that continuously aggregates information about communication links, users, the radio environment, mobility, and network resources. Unlike the domain-specific DTs, which typically hold state for a single network entity, this knowledge base is designed to persist and accumulate across tasks, so knowledge learned while solving one problem, for example a beam prediction task, remains available and useful when the network later needs to reason about mobility management or service placement.

Two structural choices distinguish this knowledge base from a conventional synchronized replica. First, spatial information, including the radio environment and XL-MIMO channel state, is organized hierarchically, ranging from coarse array-level regions to fine-grained visibility regions, so the attention mechanism can request updates at the granularity required by a given decision, thereby avoiding unnecessary full-aperture refresh. Second, the knowledge base is partitioned into slice-aware fragments, one for each active network slice such as URLLC, eMBB, or mMTC. Each fragment carries its own freshness target derived from the slice's service level agreement, so a low-latency fragment is kept close to real time while a delay-tolerant fragment is allowed to age between updates. The knowledge base also absorbs information from the users and network-services domains, encoding service requirements and operator objectives that guide subsequent reasoning. This persistence, combined with hierarchical and slice-aware organization, transforms the DT from a task-specific optimization tool into a standing, differentiated memory that PRISM can query at the resolution and freshness required by each decision.

\subsection{Intelligent Reasoning and Decision-Making: Predictive and Slice-Aware}

The reasoning module interprets the current state of the knowledge base and produces decisions that account for interactions across domains rather than optimizing a single function in isolation. Drawing on computing and AI resources, this module can weigh trade-offs such as the one between beamforming gain and sensing quality, or between aggressive mobility prediction and conservative resource reservation, and select actions that PRISM can query at the resolution and freshness required by each decision.

What makes this reasoning predictive rather than purely reactive is that it does not wait for a decision to become urgent before acting on it. By tracking how quickly each slice fragment in the knowledge base is drifting toward its service level boundary, for example a rising handover probability for a high-mobility user on a low-latency slice, the reasoning module can identify decisions it is likely to face in the next few sensing cycles and issue the attention queries described earlier ahead of time. This is also where operator intent enters the loop, since service-level objectives and policies from the network-services domain are treated as inputs that shape which anticipated decisions receive priority attention, rather than as external constraints applied after an already-optimized network state has been produced. Since reasoning draws on persistent, multi-domain, slice-aware knowledge instead of a narrow, task-specific snapshot, it can also identify dependencies that a domain-specific optimizer would not see, for instance recognizing that a beamforming decision made to serve one user on an eMBB slice degrades sensing coverage needed by a URLLC slice in the same region.

\subsection{Autonomous Planning and Continual Learning: Anticipatory Action Staging}

The final block turns reasoning outputs into concrete control actions and closes the loop back to the physical network. This module handles autonomous planning, translating decisions from the reasoning module into actions across mobility management, network services, and communication configuration, while a continual-learning component tracks how well previous actions performed and updates the models used throughout the engine accordingly.

Since reasoning already anticipates likely decisions before they become urgent, planning does not need to start from scratch when a triggering condition arises. Instead, the planning module pre-computes and stages a small set of candidate actions, while the corresponding decision is still being formed. Examples include shortlisting candidate beam pairs for a user approaching a coverage boundary or pre-reserving resources for a network slice nearing its latency threshold. When the triggering condition is finally confirmed, the network only needs to commit one of the pre-staged actions rather than run the full perception, knowledge, and reasoning sequence from scratch, which collapses the critical path latency down to little more than the actuation step itself. This closes the perception-knowledge-reasoning loop at the center of the architecture: planning decisions change the physical network, the perception module observes the outcome, the knowledge base is updated, and reasoning adjusts its next decision and its next round of anticipated decisions based on this refreshed picture. Over time, continual learning improves the accuracy of these anticipations, so the models used to predict which decisions are coming, and which parts of the knowledge base will need attention, keep improving as the network encounters more traffic patterns, mobility behavior, and slice demand profiles.

\subsection{Benefits for Ultra-Low Latency, Bandwidth Efficiency, XL-MIMO, and Network Slicing}

The four design choices described above were not added as independent features. They compound to target four requirements that conventional passive, environment-centric DT-assisted ISAC frameworks struggle to meet at the same time.

\begin{itemize}
\item \textbf{Ultra-low latency.} PRISM anticipates upcoming decisions and pre-stages candidate actions, avoiding a complete sense--reason--act cycle when a decision becomes urgent. This reduces response latency for URLLC and time-critical XL-MIMO beam realignment.

\item \textbf{Bandwidth efficiency.} PRISM refreshes only the regions and knowledge fragments relevant to active or anticipated decisions, while exchanging compact updates instead of raw measurements. This reduces signaling and fronthaul overhead compared with continuously synchronized DTs.

\item \textbf{XL-MIMO scalability.} PRISM organizes channel knowledge by subarray visibility regions and refreshes only those relevant to current decisions. This avoids repeated full-aperture estimation and keeps sensing overhead manageable as array size increases.

\item \textbf{Network slicing.} PRISM maintains slice-specific knowledge fragments with freshness targets aligned with their service-level requirements. This prioritizes near-real-time updates for latency-critical slices while conserving sensing and communication resources for delay-tolerant ones.
\end{itemize}

\begin{figure*}[t]
\centering
\captionsetup[subfigure]{
    font={footnotesize},
    labelfont=bf,
    justification=centering,
    skip=3pt
}

\begin{subfigure}[b]{0.6\textwidth}
    \centering
    \fbox{\includegraphics[width=1.2\linewidth,
                           height=3.7cm,
                           keepaspectratio]{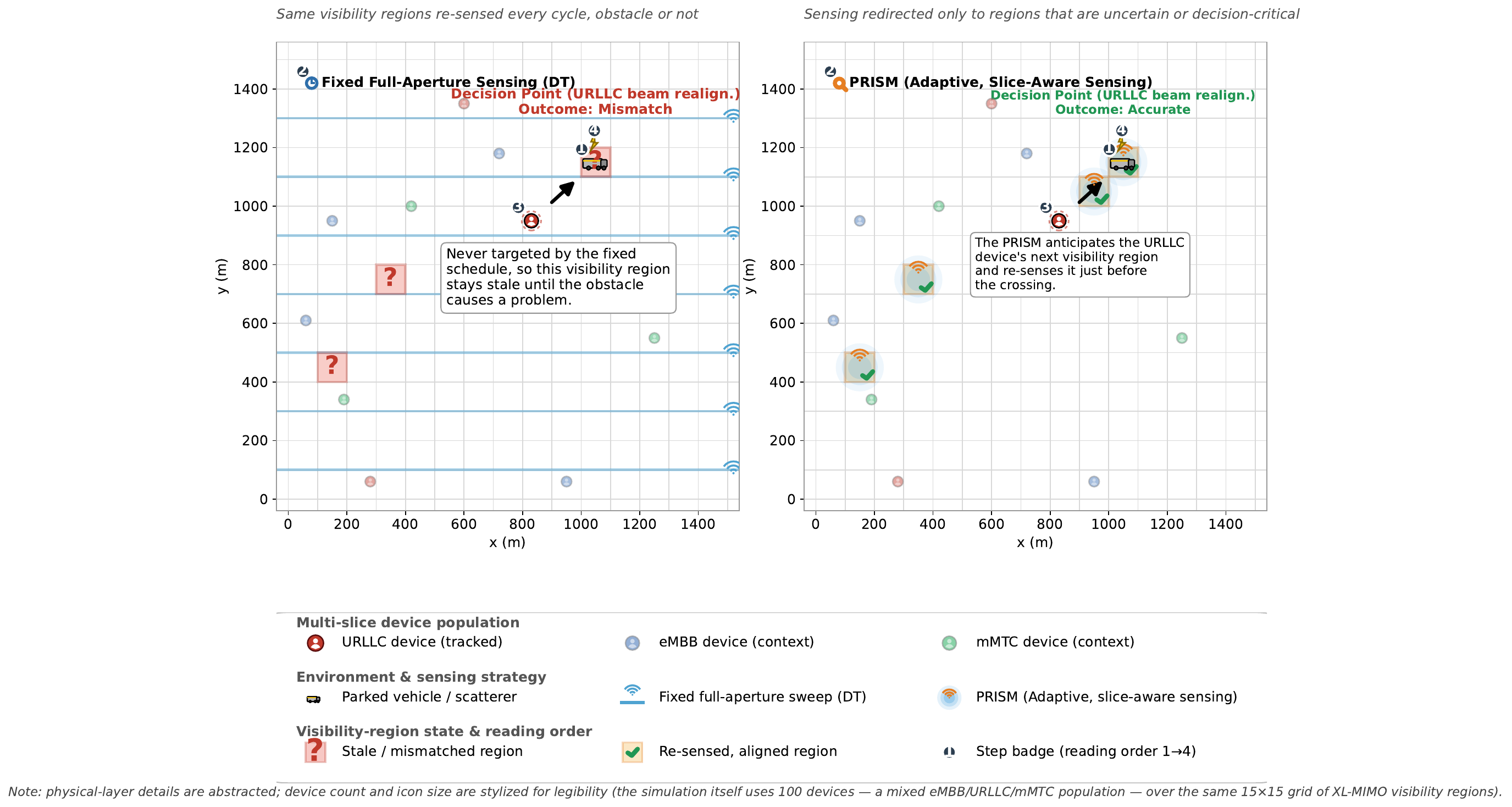}}
    \caption{}
    \label{fig3a}
\end{subfigure}
\hfill
\begin{subfigure}[b]{0.32\textwidth}
    \centering
    \fbox{\includegraphics[width=0.98\linewidth,
                           height=3.7cm,
                           keepaspectratio]{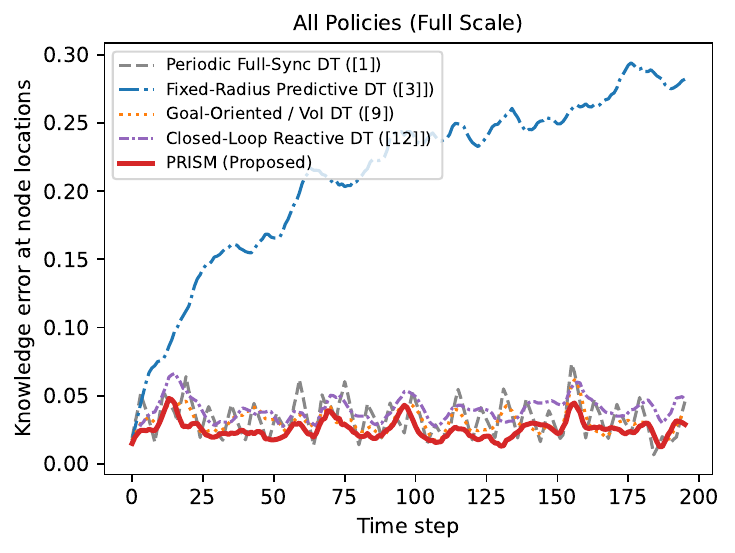}}
    \caption{}
    \label{fig3b}
\end{subfigure}

\vspace{0mm}

\begin{subfigure}[b]{0.32\textwidth}
    \centering
    \fbox{\includegraphics[width=0.98\linewidth,
                           height=3.0cm,
                           keepaspectratio]{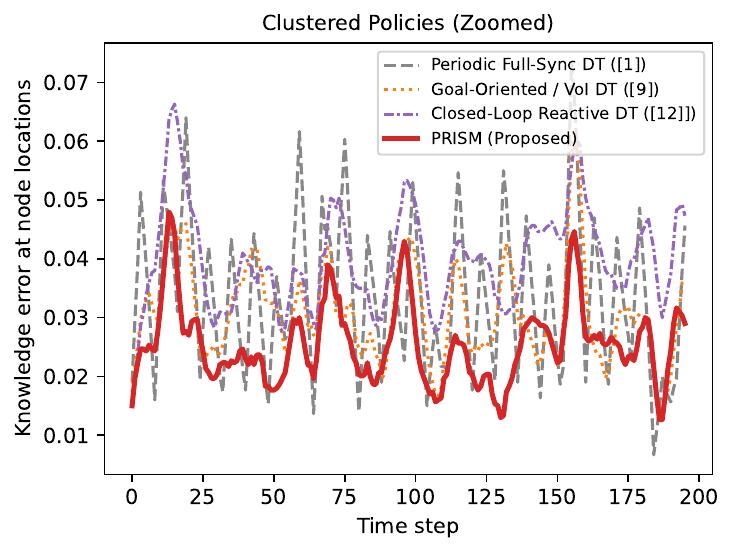}}
    \caption{}
    \label{fig3c}
\end{subfigure}
\hfill
\begin{subfigure}[b]{0.32\textwidth}
    \centering
    \fbox{\includegraphics[width=0.98\linewidth,
                           height=3.0cm,
                           keepaspectratio]{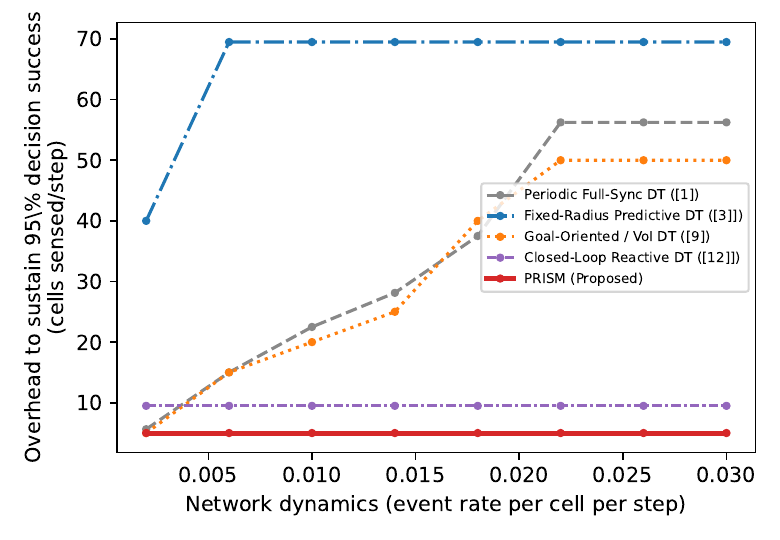}}
    \caption{}
    \label{fig3d}
\end{subfigure}
\hfill
\begin{subfigure}[b]{0.32\textwidth}
    \centering
    \fbox{\includegraphics[width=0.98\linewidth,
                           height=3.0cm,
                           keepaspectratio]{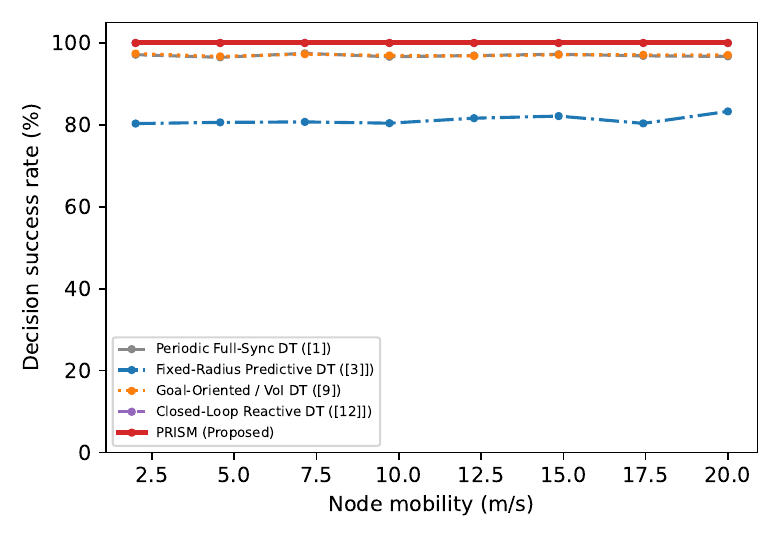}}
    \caption{}
    \label{fig3e}
\end{subfigure}

\vspace{0mm}

\caption{Preliminary evaluation of the proposed PRISM engine.
(a)~Illustrative deployment scenario.
(b)~Knowledge error at device locations across the array's visibility regions.
(c)~Zoom on the four closely clustered policies, excluding the fixed-radius outlier.
(d)~Sensing overhead needed to sustain a 95\% decision success rate as scatterer and blockage dynamics increase.
(e)~Decision success rate versus device mobility at a fixed sensing budget.}

\label{fig:energyeval}

\end{figure*}

Section~\ref{secIV} provides preliminary use case evidence for the latency and bandwidth-efficiency benefits above, through the sensing overhead and decision latency results in Fig.~\ref{fig3d} and Fig.~\ref{fig3e}. The XL-MIMO and network-slicing benefits are argued from the architecture in this section and tested only at the level of visibility-region sensing overhead in that same use case, since a full channel-level and slice-differentiated evaluation is left for future work.

\subsection{Toward Practical Deployment}

PRISM is conceived as a logical architecture rather than being bound to a fixed location, enabling its functions to be flexibly  distributed according to their latency and computational requirements. Network-wide knowledge aggregation, long-term reasoning, policy handling, and model learning can reside at regional edge or cloud infrastructure, while latency-critical perception and control should remain closer to the radio access network (RAN). Such a hierarchical deployment also reduces the transport of raw ISAC measurements, since local perception can convert them into structured knowledge updates before they traverse the network.

Open-RAN (O-RAN) provides one possible realization of this separation. Longer-timescale knowledge management, policy-driven reasoning, and learning could be implemented through RAN Applications associated with the Non-Real-Time RAN Intelligent Controller (Non-RT RIC), while latency-sensitive attention queries and control actions could be supported through eXtended Applications (xApps) in the Near-Real-Time RIC. Operator policies and service-level objectives can consequently enter the reasoning loop from the management layer, while resulting actions are executed through existing RAN control interfaces. PRISM is nevertheless independent of O-RAN, similar functional placement could be realized through multi-access edge computing or operator-specific cloud-native architectures.

\begin{table}[!t]
\caption{Benchmark policies used in the use case, mapped to their architectural counterparts in Table~\ref{tab:comparison}.}
\label{tab:usecase-benchmarks}
\centering
\scriptsize
\renewcommand{\arraystretch}{1.05}
\begin{tabular}{|p{2.7cm}|p{2.7cm}|p{1.4cm}|}
\hline
\rowcolor{bluecol}
\textbf{Architectural Category} & \textbf{Simulated Policy} & \textbf{Reference} \\
\hline
Periodic full-sync DT & Fixed-schedule, full-grid sensing & \cite{wei_integrated_2024} \\
\hline
Fixed-radius predictive DT & Mobility lookahead, no uncertainty awareness & \cite{ding_joint_2024} \\
\hline
Goal-oriented / value-of-information (VoI) DT & Senses the most uncertain cells each step & \cite{saggese_goal-oriented_2026} \\
\hline
Closed-loop reactive DT & Senses on demand exactly when a decision occurs & \cite{luo_electromagnetic_2026} \\
\hline
\end{tabular}
\end{table}

\begin{table}[!t]
\caption{Preliminary simulation results at a shared operating point (moderate mobility, moderate event rate).}
\label{tab:usecase-summary}
\centering
\scriptsize
\renewcommand{\arraystretch}{1.05}
\begin{tabular}{|p{2.4cm}|p{1.0cm}|p{1.0cm}|p{1.0cm}|p{1.0cm}|}
\hline
\rowcolor{bluecol}
\textbf{Policy} & \textbf{Know. Err.} & \textbf{Overhead} & \textbf{Success (\%)} & \textbf{Latency} \\
\hline
Periodic full-sync~\cite{wei_integrated_2024} & 0.035 & 28.5 & 96.7 & 0.000 \\
\hline
Fixed-radius predictive~\cite{ding_joint_2024} & 0.238 & 30.0 & 76.2 & 0.000 \\
\hline
Goal-oriented / VoI~\cite{saggese_goal-oriented_2026} & 0.033 & 30.0 & 96.8 & 0.000 \\
\hline
Closed-loop reactive~\cite{luo_electromagnetic_2026} & 0.042 & 9.7 & 100.0 & 1.000 \\
\hline
\textbf{Proposed PRISM} & \textbf{0.026} & 30.0 & \textbf{100.0} & \textbf{0.009} \\
\hline
\end{tabular}
\end{table}

\section{Use Case: PRISM Engine for XL-MIMO Sensing and Multi-Slice Operation in a Dense Urban 6G Deployment}
\label{secIV}
 
\subsection{Scenario and Simulation Methodology}
 
The numerical study here validates decision-centric predictive perception, the adaptive sensing and knowledge-update mechanism at the core of PRISM, rather than the complete architecture. We consider a dense urban 6G deployment served by an XL-MIMO base station that multiplexes eMBB, URLLC, and mMTC traffic on shared radio resources. An unknown obstacle, for example a parked vehicle or a new scatterer, perturbs the near-field propagation seen by part of the array without warning. This scenario stress-tests decision-centric predictive perception, since it combines localized, unpredictable environmental change with continuously moving devices drawn from a mixed-slice population. It is depicted in Fig.~\ref{fig3a}
 
We build a lightweight system-level simulation that captures the information flow the architecture is designed around. The coverage area is a $15 \times 15$ grid of cells over a $1500 \times 1500$ meter urban region, each cell representing a visibility region of the XL-MIMO array. The base station serves 100 devices representing eMBB, URLLC, and mMTC services, with device mobility modeled using the random-waypoint model. Unknown events occur at a configurable rate, flipping a visibility region's true state and creating a mismatch with the DT's last known estimate until that region is sensed again, representing a new scatterer or blockage entering the near-field environment. Every time a device crosses into a new cell, this is treated as a decision point, analogous to a beam realignment or handover decision, and is used to evaluate whether the DT's knowledge was accurate at the moment it was needed.
 
\subsection{Benchmarks and Metrics}
 
This is a system-level simulation of information flow, implemented in Python, rather than a link-level simulation, so path loss, channel models, and waveform parameters are intentionally not modeled, consistent with the scope stated above. We compare the PRISM engine against four sensing and knowledge-update policies, each constructed to represent the architectural category of a corresponding entry in Table~\ref{tab:comparison} rather than re-implementing its specific algorithm. Table~\ref{tab:usecase-benchmarks} summarizes this mapping. For each policy, including the PRISM engine, we record four metrics: knowledge error at device locations, which measures visibility-region error where it matters for XL-MIMO channel acquisition rather than across the whole array; sensing overhead, the average number of visibility regions sensed per step, a proxy for pilot and signaling load; decision success rate, the fraction of slice-relevant decisions made with already-correct knowledge; and decision latency, the extra steps a policy spends re-sensing before a decision can be finalized, most consequential for URLLC reliability.
 
\subsection{Results}
 
Fig.~\ref{fig3b} tracks knowledge error at device locations over time at a moderate mobility and event-rate operating point. It shows that a fixed-radius predictive baseline with no uncertainty awareness drifts far from the true state as devices continue moving, since it keeps sensing wherever devices are heading regardless of whether that information is still needed. Fig.~\ref{fig3c} zooms in on the remaining four policies, which move in a visibly tighter band. Averaged over six random seeds, the PRISM engine reaches a mean knowledge error of 0.026, compared with 0.033 for the goal-oriented baseline, 0.035 for periodic full-sync, 0.042 for closed-loop reactive, and 0.238 for fixed-radius prediction. Lower, more consistent knowledge error at device locations is what accurate XL-MIMO beam and CSI acquisition requires, since it is the local, per-device channel view that determines beamforming quality.

Fig.~\ref{fig3d} depicts how much sensing overhead each policy needs to sustain a $95\%$ percent decision success rate as scatterer and blockage dynamics increase. This is where decision-centric predictive perception pays off most clearly, and it is the result closest to the bandwidth efficiency and XL-MIMO scalability benefits. The PRISM engine sustains the target with a flat overhead of about five visibility regions per step across the entire tested range, since the regions it senses are tied to imminent device transitions rather than to how often the near-field environment changes. The periodic and goal-oriented baselines must instead increase overhead more than tenfold, from roughly $5$ to $6$ regions per step at low dynamics up to $50$ and $56$ at the highest tested event rate, while the fixed-radius baseline never becomes efficient at any dynamics level. The closed-loop reactive baseline stays flat at a low overhead of about $10$ regions per step, since it only senses exactly where a decision is happening, but as the next figure shows, this comes at a latency cost the proposed architecture avoids.

Fig.~\ref{fig3e} reports decision success rate as device mobility increases, at a fixed sensing budget shared by the budget-limited policies. The PRISM engine and the closed-loop reactive baseline both sustain a full $100\%$ success rate across the tested range, the periodic and goal-oriented baselines settle around $96\%$ to $97\%$ regardless of speed, and the fixed-radius baseline stays substantially degraded near $80\%$ throughout. This figure alone omits that the proposed engine matches the reactive baseline's success rate at a decision latency of 0.009 steps against 1.0 steps, roughly two orders of magnitude lower for the same reliability, since most decisions are already resolved before needed, the gap that matters most for millisecond-scale URLLC decisions. Table~\ref{tab:usecase-summary} summarizes all four metrics at the shared operating point used for Fig.~\ref{fig3b}, averaged over multiple seeds.

\section{Conclusion and Future Directions}

This article introduced a Predictive, Reasoning-driven, Intelligent Sensing Module (PRISM) engine, reshaping ISAC-assisted DTs from passive, domain-specific optimizers into a persistent, network-wide reasoning system built around decision-centric predictive perception. Sensing is directed toward anticipated decisions instead of a fixed schedule, targeting ultra-low latency, bandwidth-efficient sensing, XL-MIMO channel acquisition, and slice-aware operation across eMBB, URLLC, and mMTC services. Preliminary simulations show large overhead reductions and near-zero decision latency versus representative baselines. Future work includes link-level evaluation with realistic near-field models, slice-differentiated scheduling, and hardware validation toward self-aware 6G networks.

\bibliographystyle{IEEEtran}
\bibliography{Ref}

\end{document}